
\magnification=\magstep1
\baselineskip=24 true pt
\hsize=33 pc
\vsize=44 pc
\def \clg {{\cal G}}
\centerline {\bf STRING EFFECTIVE ACTION AND}
\centerline {\bf TWO DIMENSIONAL CHARGED BLACK HOLE }
\bigskip
\bigskip
\bigskip
\centerline {\bf S. Pratik Khastgir and Alok Kumar}
\bigskip
\centerline {\it { Institute of Physics}}
\centerline {\it { Bhubaneswar-751005}}
\centerline {\it { INDIA}}
\bigskip
\bigskip
\bigskip
\bigskip
\centerline {\bf Abstract}

Graviton-dilaton background field equations in threee space-time dimensions,
 following from the string
 effective action are solved when the metric
 has only time dependence. By taking one of the two space
dimensions as compact, our solution reproduces a recentlty discovered
charged black hole solution in two space-time dimensions. Solutions
in presence of nonvanishing three dimensional background
antisymmetric tensor field are also discussed.
\vfil
\eject

Black hole solutions in string theory has recently been found both as
an exact conformal field theory solution$^{1,2}$ and
 as a solution of the field
equations in effective field theory$^{3}$ of strings
. Exact conformal field theory
solution is obtained as an SL(2,R)/U(1) coset conformal field theory,
formulated as a gauged Wess-Zumino-Witten(WZW) model for SL(2,R)
group with gauging of its U(1) subgroup. In the
effective field theory approach, these solutions are characterized by
the nontrivial classical values for the metric and dilaton fields which
satisfy the background field equations.

Recently, a generalization of these results was done by finiding the
two dimensional charged black hole solution$^{4}$ as an
 [SL(2,R)$\times$U(1)${_i}$]/U(1)
coset conformal field theory, where the group U(1)$_{i}$
 corresponds to an internal compact
dimension. This solution is also obtained as a gauged WZW model. The
gauged U(1) subgroup now is a combination of
 U(1)${_i}$ and a U(1) subgroup
of SL(2,R).

In a related development recently, Meissner and Veneziano$^{5}$
 wrote down the field equations for the D-dimensional
   superstring effective
action in a general time dependent graviton,
 dilaton and antisymmetric tensor
background and showed its invariance under an O(d,d)
 group of transformations, where d is the number of space dimensions (d=D-1).
In ref. [6] these equations were written down and solved for
arbitrary D, when dilaton and diagonal components of the metric
 are present as background. For
D=2 these solutions are identical to the black hole solutions$^{5}$.

In this paper we solve the field equations of Meissner and
 Veneziano$^{5}$
in three dimensions when metric has off diagonal components as
 well. After compactification of one of the space dimensions, our
solution can be interpreted
 as two dimensional charged black hole$^{4}$. We also show that, by
making O(2,2) transformations, one can obtain solutions with
nonvanishing background antisymmetric tensor field.

Our starting point is the genus zero low energy
 effective action for closed
 superstrings in the limit when string tension $\alpha '\rightarrow 0 $.
 Restricting to the graviton, dilaton and  antisymmetric tensor
field, this action in D-dimensions is written as,
$$ S=\int d^{D}x \sqrt{-detG}
e^{-\phi}\big [V -R-G^{\mu\nu}\partial_{\mu}\phi\partial_{\nu}\phi-{1\over
12}H_{\mu\nu\rho}H^{\mu\nu\rho}\big ]\eqno(1)$$

\noindent where $V$ contains cosmological constant as well as the dilaton
 potential, which for our solution turns out to be a constant. $\phi$ is the
dilaton field, $G_{\mu\nu}$ is the D-dimensional metric and $H_{\mu\nu\rho}$
is the field strength for the antisymmetric tensor field $B_{\mu\nu}$:

$$H_{\mu\nu\rho}=\partial_{\mu}B_{\nu\rho}+cyclic. \eqno(2)$$

We consider the case when D=3. In general, when D is less than the critical
dimension of (super)strings, massless gauge and Higgs fields can also
be present. However we consider the case when they have vanishing
background values.

As in refs.[5,6], we now look for the solutions of the field equations
when $G$ and $B$ are functions of time only. In this case, gauge
symmetries of the action, eqn. (1), allow $G$ and $B$ to be always brought
in the form$^{5}$:

$$G=\bigg (\matrix {{-1} & {0}\cr {0} &{ \clg (t)} \cr }\bigg ),
\qquad   B=\bigg (\matrix
{{0} & {0} \cr {0} & {{\cal B} (t)} \cr}\bigg ).\eqno(3)$$

\noindent  Then, action (1) can be rewritten as
$$ S=\int dt \sqrt{det\clg}
e^{-\phi}\big [V -2\partial_{0}^{2}(ln\sqrt {det \clg})-
(\partial_{0}(ln\sqrt {det \clg}))
^{2}+{1 \over 4}Tr(\partial_{0} \clg)(\partial_{0} \clg^{-1})
+(\partial_{0} \phi)^{2}$$$$ +{1\over 4}Tr(\clg^{-1}(\partial_{0}{\cal B})
{\clg^{-1}(\partial_{0}{\cal B}))}\big ].\eqno(4)$$

\noindent We will first consider the case of vanishing background $B$ field,
and discuss later on the case when it has nonzero value.
By redefining the dilaton field :
$$\Phi=\phi-ln\sqrt{det \clg}, \eqno(5)$$
\noindent one can then write eqn. (4) as:

$$ S=\int dt
e^{-\Phi}[V +\dot \Phi^{2}-{1\over 4}Tr(\clg^{-1} \dot
\clg)^{2}].\eqno(6)$$

Equations of motion for these fields$^{5}$ are:
$$(\dot \Phi)^{2}-{1\over 4}Tr[(\clg^{-1} \dot
\clg)(\clg^{-1} \dot
\clg)]-V =0 \eqno(7)$$

$$(\dot \Phi)^{2}-2\ddot \Phi +{1\over 4}Tr[(\clg^{-1} \dot
\clg)(\clg^{-1} \dot
\clg)]-V =0 \eqno(8)$$

\noindent and
$$-\dot \Phi \dot \clg +\ddot \clg -\dot \clg \clg^{-1}\dot \clg =0.\eqno(9) $$

\noindent where eqns. (8) and (9) follow from
 the variations with respect to
the fields $\Phi$ and $\clg_{ij}$ of action (6). Equation (7), which
is obtained directly from the variation of the action (1) with
respect to $G_{00}$, is also called the "zero energy condition"$^{5}$. It has
also been pointed out in ref. [5] that eqns. (7)-(9) are the
complete set of equations of motion. In ref. [6] these equations were
 solved for arbitrary D when $\clg$ is diagonal. For D=2 the solution
  is identical to the uncharged black hole solution when roles of space
and time are interchanged.

We now obtain a solution of eqns.(7)-(9) when $\clg$ is a $2 \times 2$
symmetric matrix of the form:

$$\clg (t)=\bigg (\matrix{g_{1}(t) & {1\over 2}A(t)\cr {1\over 2}A(t) &
g_{2} (t) \cr }\bigg )\eqno(10)$$

In general, in terms of these components, eqns. (7)-(9) will be
 very complicated.
To get these equations in a rather convenient form, we now redefine:

$$g_{1}=\tilde g_{1} +{1\over 4}{A^{2}\over g_{2}}.\eqno(11)$$

\noindent Above form of the metric has also been used for constant
 backgrounds in ref. [7]. The determinent of $\clg$ is then simply,

$$det \clg =\tilde g_{1} g_{2}, \eqno(12)$$

\noindent the inverse metric is given as,

$$\clg^{-1}={1\over \tilde g_{1} g_{2}}\bigg (\matrix{g_{2} &
-{1\over 2}A\cr -{1\over 2}A &
g_{1} \cr }\bigg ),\eqno(13)$$

\noindent and $\dot \Phi=\dot \phi-{1\over 2}{\dot {\tilde g_{1}}\over \tilde
g_{1}}-{1\over 2}{\dot g_{2}\over g_{2}}$.
 In component form, eqns. (7)-(9) can therefore be written as,
$$(\dot \Phi)^{2} - {1\over 4}[{\dot {\tilde g_{1}^{2}}\over \tilde
g_{1}^{2}} +{1\over 2}{\dot A^{2}\over \tilde g_{1}g_{2}}-
{A\dot A \dot g_{2}\over
\tilde g_{1} g_{2}^{2}}+{1\over 2}{A^{2}\dot g_{2}^{2}\over \tilde
g_{1}g_{2}^{3}} +{\dot g_{2}^{2}\over g_{2}^{2}}]-V = 0\eqno(14)$$
$$(\dot \Phi)^{2} - 2\ddot \Phi+ {1\over 4}[{\dot {\tilde g_{1}^{2}}\over
 \tilde g_{1}^{2}} +{1\over 2}{\dot A^{2}\over \tilde g_{1}g_{2}}
-{A\dot A \dot g_{2}\over
\tilde g_{1} g_{2}^{2}}+{1\over 2}{A^{2}\dot g_{2}^{2}\over \tilde
g_{1}g_{2}^{3}} +{\dot g_{2}^{2}\over g_{2}^{2}}]-V = 0\eqno(15)$$

$$-\dot \Phi \dot g_{1}+\ddot g_{1}-{1\over \tilde
g_{1}g_{2}}(g_{2}\dot g_{1}^{2}-{1\over 2}A \dot A\dot g_{1}+{1\over
4}g_{1} \dot A^{2})=0 \eqno(16)$$
$$-\dot \Phi \dot g_{2}+\ddot g_{2}-{1\over \tilde
g_{1}g_{2}}(g_{1}\dot g_{2}^{2}-{1\over 2}A \dot A\dot g_{2}+{1\over
4}g_{2} \dot A^{2})=0 \eqno(17)$$
$$-\dot \Phi \dot A+\ddot A-{1\over \tilde
g_{1}g_{2}}(g_{2}\dot A\dot g_{1}-A \dot g_{1}\dot g_{2}+g_{1}\dot A
\dot g_{2}-{1\over
4}A \dot A^{2})=0 \eqno(18)$$
\noindent where eqns. (14) and (15) follow from eqns. (7) and (8) respectively,
 and eqns. (16)-(18) are the component form of eqn. (9). We now show
that eqns. (14)-(18) have a solution of the following form :
 $$g_{1}=a_{0}+b_{0} tanh^{2}t \eqno(19)$$
$$A=b_{1} tanh^{2}t \eqno(20)$$
$$g_{2}=b_{2} tanh^{2}t \eqno(21)$$
\noindent and
$$\phi = -log(cosh^{2}t) + a \eqno(22)$$
\noindent with
$$4b_{0}b_{2}=b_{1}^{2}. \eqno (23)$$
\noindent Putting eqns. (19)-(23) into eqn. (11), we get $\tilde g_{1}=
a_{0}$, which is a constant. For the above choice of solution,
 eqns. (16)-(18) then imply:
$$(4b_{0}-{b_{1}^{2}\over b_{2}}){1\over cosh^{4}t}-({b_{0}\over
a_{0}b_{2}})(4b_{2}b_{0}-b_{1}^{2}){tanh^{2}t\over cosh^{4}t}=0 \eqno(24)$$
$$-{1\over a_{0}}(4b_{2}b_{0}-b_{1}^{2}){tanh^{2}t\over cosh^{4}t}=0
 \eqno(25)$$
\noindent and
$$(4b_{2}b_{0}-b_{1}^{2}){tanh^{2}t\over cosh^{4}t}=0
 \eqno(26)$$
\noindent and are therefore satisfied due to the condition (23).
Similarly, since
$$\dot \Phi^{2}= (\dot \phi-{1\over 2}{\dot {\tilde g_{1}}\over
\tilde g_{1}} -{1\over 2}{\dot g_{2}\over g_{2}})^{2}={1\over
sinh^{2}tcosh^{2}t} +4  \eqno(27)$$
$$\ddot \Phi=(\ddot \phi-{1\over 2}{\ddot {\tilde g_{1}}\over
\tilde g_{1}} -{1\over 2}{\ddot g_{2}\over g_{2}} +{1\over 2}{ \dot
{\tilde g_{1}^{2}}\over
\tilde g_{1}^{2}} +{1\over 2}{\dot g_{2}^{2}\over g_{2}^{2}})={1\over
sinh^{2}tcosh^{2}t}   \eqno(28)$$
\noindent and
$${1\over 4}Tr(\clg^{-1}\dot \clg)^{2}= {1\over 4}[{\dot {\tilde g_{1}^{2}}
\over \tilde
g_{1}^{2}} +{1\over 2}{\dot A^{2}\over \tilde g_{1}g_{2}}-{A\dot A \dot
g_{2}\over
\tilde g_{1} g_{2}^{2}}+{1\over 2}{A^{2}\dot g_{2}^{2}\over \tilde
g_{1}g_{2}^{3}} +{\dot g_{2}^{2}\over g_{2}^{2}}]={1\over sinh^{2}t
 cosh^{2}t}\eqno(29)$$
\noindent therefore eqns.(14) and (15) are also satisfied for $V
=4$. This value of $V$ is same as when $g_2$ and $\phi$ are the only
nonzero background fields. $g_1$ and $A$ fields therefore do not
contribute to $V$.

Our solution, eqns. (19)-(23), appear to have four independent
parameters. However, since equations of motion (14)-(18) have the
following scaling symmetry : $g_{2}\rightarrow \alpha g_{2}$,
$g_{1}\rightarrow {\alpha^{-1}} g_{1}$, $\tilde g_{1}\rightarrow
{\alpha^{-1}}
{\tilde g_{1}}$, $A\rightarrow A $ and $\Phi\rightarrow\Phi$, hence
one of the parameters can be scaled away. It will be later on seen
that the number of parameters for our solution is same as that of the
charged black hole of ref. [4].

Now, to discuss the charged black hole interpretation of our solution,
eqns. (19)-(23),let us consider the world sheet string action for a
three dimensional target space, written in presence of background
metric, gauge and Higgs fields as,
$$I=\int d^{2}z[\sum_{i,j=0,2} G_{ij}\partial X^{i}\bar\partial
X^{j}+ \sum_{i=0,2}{1\over 2}A_{i}(\partial X^{i}\bar \partial
X^{1}+\partial X^{1}\bar\partial X^{i})+\psi \partial X^{1}\bar
\partial X^{1}]\eqno(30)$$
\noindent where indices (0,2) denote the space-time index and $X^{1}$
 is a compact direction. Comparing our solution, eqns. (19)-(23)
  with eqn. (30), we get a background space-time metric of the form,
$$G_{ij}=\bigg ( \matrix {{-1} & {0 }\cr {0} &
 {b_{2} tanh^{2}t} \cr} \Bigg ),\eqno(31)$$
\noindent background gauge field:
$$A_{i}=(0,b_{1} tanh^{2}t),\eqno(32)$$
\noindent and scalar field:
$$\psi=a_{0}+b_{0} tanh^{2}t=(a_{0}+b_{0})-{b_{0}\over cosh^{2}t}.\eqno(33)$$
\noindent In addition, there is also a background
dilaton:$$\phi=-log(cosh^{2}t) +a.\eqno(34)$$
Our solution can therefore be identified with the charged black hole
solution of ref.[4], if  roles of space and time are interchanged and
 following identifications are also made:
$$b_{0}={e^{2}\over 4k^{2}}$$
$$b_{1}=-{e\over k}   \eqno(35)$$
$$b_{2}=1$$
\noindent and
$$a_{0}=-{1\over 2k}-{e^{2}\over 4k^{2}}$$
It can also be verified that the equations of motion (14)-(18) have
another solution when two dimensional metric ($\hat G_{ij}$), gauge
field ($\hat A_{i}$), Higgs ($\hat \psi$) and dilaton ($\hat \phi$)
have the following form:
$$\hat G_{ij}=\bigg ( \matrix {{-1} & {0 }\cr {0} &
{{\hat b_{2}} coth^{2}t} \cr} \Bigg ).\eqno(36)$$
$$\hat A_{i}=(0,{\hat b_{1}} coth^{2}t) \eqno(37)$$
$$\hat \psi = {\hat a_{0}+\hat b_{0} coth^{2}t}\eqno(38)$$
$$\hat \phi=-log(sinh^{2}t) +\hat a.\eqno(39)$$
\noindent and coefficients $\hat b_{i}$'s satisfy $4\hat b_{0}\hat
b_{2}= \hat b_{1}^{2}$. We note that the above solution is the "dual"
black hole solution of refs. [1,2,3] with nontrivial gauge and scalar
background.

So far we have discussed solutions which correspond to the case when
$B$ field is set to zero in the three dimensional effective action (1).
 However, it has been pointed out before in refs. [5,8,9], that nonzero $B$
field can be generated by using an O(d,d) symmetry of the effctive
action as well as the equations of motion. This O(d,d)
transformation acts as$^{5}$ :
$$M^{'}=\Omega M \Omega^{T}\eqno(40)$$
\noindent where
$$M \equiv\bigg ( \matrix {{\clg^{-1}} & {-{\clg^{-1}{\cal B}} }\cr
{{\cal B}{\clg^{-1}}} &
{\clg-{\cal B}{\clg^{-1}}{\cal B}} \cr} \Bigg ),\eqno(41)$$
\noindent and $\Omega$ is a 2d$\times$2d matrix satisfying
$$\Omega^{T}\eta\Omega=\eta\eqno(42)$$
\noindent with
$$\eta =\bigg ( \matrix {{0} & {I}\cr {I} & {0} \cr} \Bigg ).\eqno(43)$$
We now apply the above procedure to generate nonzero "$B$" field from
our solution. $B$ being zero in our original solution, $M$
is of the form:
$$M=\bigg ( \matrix {{\clg^{-1}} & {0}\cr
{0} & {\clg} \cr} \Bigg ).\eqno(44)$$
\noindent By making an O(2,2) transformation with $\Omega$:
$$\Omega =\bigg ( \matrix {{\Pi} & {1-\Pi}\cr {1-\Pi} & {\Pi} \cr} \Bigg )
\eqno(45)$$
\noindent and
$$\Pi =\bigg ( \matrix {{1} & {0}\cr {0} & {0} \cr} \Bigg ),\eqno(46)$$
\noindent one obtains,
$$M^{'} \equiv \bigg ( \matrix {{{\clg'}^{-1}} & {-{{\clg'}^{-1}{\cal B}'} }
\cr {{\cal B}'{{\clg'}^{-1}}} &
{\clg'-{\cal B}'{{\clg'}^{-1}}{\cal B}'} \cr} \Bigg ),\eqno(47)$$
\noindent where
$$\clg' =\bigg ( \matrix {{\tilde g_{1}} & {0}\cr {0} & {{1\over g_{2}}}
 \cr} \Bigg ),\eqno(48)$$
\noindent and
$${\cal B}' =\bigg ( \matrix {{0} & {{1\over 2}{A\over g_{2}}}\cr
 {-{1\over 2}{A\over g_{2}}} & {0} \cr} \Bigg ).\eqno(49)$$
\noindent Also, since under O(d,d) $\Phi\rightarrow\Phi$, hence
new dilaton field is given as
$$\phi^{'}=\phi-ln g_{2}+const.\eqno(50)$$
It is interesting to note that given our earlier solution, eqn. (10)
 and (19)-(23), $\clg'$
 and $\cal B'$ also describe a black hole solution. The gauge field
coupling in eqn. (30) is now antisymmetric. One can generate several
solutions in the above manner.

Finally, one can also try to give an alternative interpretation of our
solution when roles of the string co-ordinates $X^{1}$ and $X^{2}$,
in eqn. (30) are interchanged, so that $X^{2}$ is now the compact direction.
 In this case the
two dimensional space-time metric, now specified by the indices (0,1),
 takes the form:
$$\tilde G_{ij}=\bigg ( \matrix {{-1} & {0 }\cr {0} &
 {a_{0}+b_{0}tanh^{2}t} \cr} \Bigg ).\eqno(51)$$

We now note that, unlike the previous case, eqn.(31), the new metric
 (51)
and dilaton (34) are not a solution of the two dimensional
graviton-dilaton equation of motion for $a_{0},b_{0}\not= 0$. Only
after background gauge field
$$\tilde A_{i}=(0,b_{1}tanh^{2}t) \eqno(52)$$
\noindent and Higgs field
$$\tilde \psi=b_{2}tanh^{2}t \eqno(53)$$
\noindent are also added, that the equations of motion are satisfied.
It will be interesting to further analyze this solution and study the
space-time geometry. A generalization of  the results of this paper
to arbitrary number of dimensions is being investigated also.

\noindent{\bf Acknowledgement:}

We thank J. Maharana and A.M. Sengupta for useful discussions.
\vfil
\eject

\noindent {\bf References:}

\item 1. E. Witten, {\it Phys. Rev.} {\bf D44}, (1991) 314.
\item 2. R. Dijgraaf, E. Verlinde and H. Verlinde, Institute for
Advanced Study preprint, IASSNS-HEP-91/22 (1991);
E. Martinec and S. Shatasvili, Chicago
preprint,
EFI-91-22 (1991);
I. Bars, University of South California preprint, USC-91/HEP-B3
(1991); USC-91/HEP-B4 (1991);
M. Bershadsky and D. Kutasov, PUPT-1261, HUTP-91/A024.
\item 3. G. Mandal, A. Sengupta and S. Wadia, Institute for Advanced
 Study preprint, IASSNS-HEP-91/10 (1991);
S. P. de Alwis and J. Lykken, 91/198-T, COLO-HEP-258 (1991).
\item 4. N. Ishibashi, M. Li and Alan R. Steif, University of
California, Santa Barbara preprint, UCSBTH-91-28 (1991).
\item 5. K. A. Meissner and G. Veneziano, CERN Theory preprint,
CERN-TH-6138/91 (1991).
\item 6. G. veneziano, CERN Theory preprint, CERN-TH-6077/91 (1991).
\item 7. A. Giveon, E. Rabinovici and G. Veneziano, {\it Nucl. Phys.}
{\bf B322}
(1989) 167.
\item 8. A. Sen, TIFR Theory preorint, TIFR/TH/91-35 (1991).
\item 9. M. Gasperini, J. Maharana and G. Veneziano, CERN Theory
preprint, CERN-TH-6214/91 (1991).
\eject

\end